\documentclass[twocolumn,aps,amsmath,amsfonts,showpacs]{revtex4}
\usepackage{graphicx}

\newcommand{\be}{\begin{equation}}
\newcommand{\ee}{\end{equation}}
\newcommand{\bea}{\begin{eqnarray}}
\newcommand{\eea}{\end{eqnarray}}
\newcommand{\bn}{\begin{enumerate}}
\newcommand{\en}{\end{enumerate}}
\newcommand{\bi}{\begin{itemize}}
\newcommand{\ei}{\end{itemize}}

\newcommand{\ie}{\emph{i.e., }}
\newcommand{\etal}{\emph{et~al.~}}

\newcommand{\cc}{c(r,\theta,\phi,\Psi_1,\Psi_2)}
\newcommand{\ccR}{c(R,\theta,\phi,\Psi_1,\Psi_2)}

\newcommand{\dall}{\int d\Omega \int d\Psi_1 \, \int d\Psi_2 \,}
\newcommand{\kDC}{k_{\text{DC}}} 
 
\newcommand{\Darot}{D_1^{\text{rot}}} 
\newcommand{\Dbrot}{D_2^{\text{rot}}} 
\newcommand*{\sfr}{\mbox{\textbf{\textsf{r}}}}

\newcommand{\alm}{A_{ll_1l_2}^{mm_1n_1m_2n_2}}

\begin{document}

\title{A general expression for bimolecular association rates\\with
  orientational constraints}

\author{Maximilian Schlosshauer}

\email{MAXL@u.washington.edu}

\affiliation{Department of Physics, University of Washington, Seattle,
  WA 98195}

\author{David Baker} 

\email{dabaker@u.washington.edu} 

\thanks{To whom correspondence should be addressed.  Mailing address:
  Department of Biochemistry, University of Washington, Box 357350,
  Seattle, WA 98195. Telephone (206) 543-1295, Fax (206) 685-1792.}

\affiliation{Department of Biochemistry, University of Washington,
  Seattle, WA 98195}

\begin{abstract}
  We present a general expression for the association rate for
  partially diffusion-controlled reactions between spherical molecules
  with an asymmetric reactive patch on each surface. Reaction can
  occur only if the two patches are in contact and properly aligned to
  within specified angular tolerances. This extends and generalizes
  previous approaches that considered only axially symmetric patches;
  the earlier solutions are shown to be limiting cases of our general
  expression. Previous numerical results on the rate of
  protein--protein association with high steric specificity are in
  very good agreement with the value computed from our analytic
  expression. Using the new expression, we investigate the influence
  of orientational constraints on the rate constant. We find that for
  angular constraints of $\sim 5^{\text{o}}$--15$^{\text{o}}$, a
  typical range for example in the case of protein--protein
  interactions, the reaction rate is about 2 to 3 orders of magnitude
  higher than expected from a simple geometric model.  \\[.3cm] Journal
  reference: \emph{J.~Phys.~Chem.~B }\textbf{106}(46), 12079--12083
  (2002).
\end{abstract}


\maketitle

\section{Introduction}

The association of two macromolecules, in particular the formation of
protein--protein complexes, is an ubiquitous process in biology. In
the simplest case of the associating species being modeled as
uniformly reactive spheres, the diffusion-controlled association rate
is given by the classic Smoluchowski result \cite{smoluchowski17},
$\kDC=4\pi D R$, where $D$ is the relative translational diffusion
constant and $R$ denotes the sum of the radii of the molecules.
Typically, however, successful complex formation hinges on the proper
relative orientation of the reactants, which can be represented by
molecules carrying reactive surface patches that have to come into
contact with high steric specificity for the reaction to occur.

The simple approach of multiplying the Smoluchowski rate constant for
uniformly reactive molecules by the probability that in a random
encounter the two molecules are properly oriented (``geometric rate'')
yields rate constants that are commonly several orders of magnitude
lower than the observed values. Some authors attributed this puzzling
behavior to the presence of long-range attractive interactions between
the molecules that not only generally speed up the rate of encounter
of the molecules but also help ``guide'' the molecules into
configurations close to the proper mutual orientation.

In addition to this approach, various attempts have been made to
quantitatively elucidate the influence of orientational constraints
and rotational diffusion on the association rate constant. Among the
earliest studies, \v{S}olc and Stockmayer derived a formal solution
\cite{solc71} of the association rate constant of spherical molecules
with axially symmetric distributions of reactivity and presented
numerical results \cite{solc73} for the simplified case of one of the
molecules being uniformly reactive. Schmitz and Schurr
\cite{schmitz72} investigated both analytically and numerically the
problem of the reaction between mobile orientable spheres, carrying
single, axially symmetric reactive patches on their surface, with
localized hemispherical sites on a plane. Shoup \etal \cite{shoup81}
introduced a generally applicably approximative treatment that allowed
simplification of the complex formal solutions of \v{S}olc and
Stockmayer \cite{solc73} and Schmitz and Schurr \cite{schmitz72} to
closed analytical expressions; this approximation was also used by
Zhou \cite{zhou93} in deriving an expression for the association rate
when each molecule bears an axially symmetric reactive patch.  All
these approaches showed that, because of relative angular
reorientations caused by translational and rotational diffusion, the
reduction in association rate brought about by orientational
constraints is significantly less than suggested by the reduction in
the probability for a properly oriented encounter.

The previous analytical treatments, however, impose only (at most)
axially symmetric orientational constraints, whereas no analytical
treatment has been presented thus far for the general case of
asymmetric reactive patches (as in the important case of sterically
highly specific protein--protein interactions), where the precise
relative orientation of the binding partners has to be specified and
appropriately constrained.

The only numerical estimates for the association rate constant for
this general case stem from Brownian Dynamics simulations, as for
example performed by Northrup and Erickson \cite{northrup92}, who
consider diffusional association of spherical molecules,  each
bearing a reactive patch composed of four contact points in a square
arrangement on a plane tangential to the surface of the molecules;
reaction is then assumed to occur if three of the four contact points
are correctly matched and within a specified maximum distance. The
rate constants are again found to be about 2 orders of magnitude
higher than expected from a simple geometric argument, but as the
approach is not analytical, the result is not readily generalizable.

In the following, we present a general expression for the partially
diffusion-controlled rate constant $\kDC$ for two spherical molecules
with fully asymmetric binding patches. The theoretical derivation is
given in Sec.~\ref{sec:theory}. Various aspects of our general
expression are investigated in Sec.~\ref{sec:results}, where we
demonstrate that previous approaches are, as expected, limiting cases
of our general treatment (Sec.~\ref{sec:limits}), discuss the
dependence of the rate constant on orientational constraints (Section
\ref{sec:values}), and compare numerical values obtained from our
expression with the result of a Brownian Dynamics simulation by
Northrup and Erickson \cite{northrup92} (Sec.~\ref{sec:bd}).

\section{\label{sec:theory}Theory} 

\subsection{\label{sec:cs}Model and coordinate system} 

\begin{figure}
\includegraphics[scale=0.75]{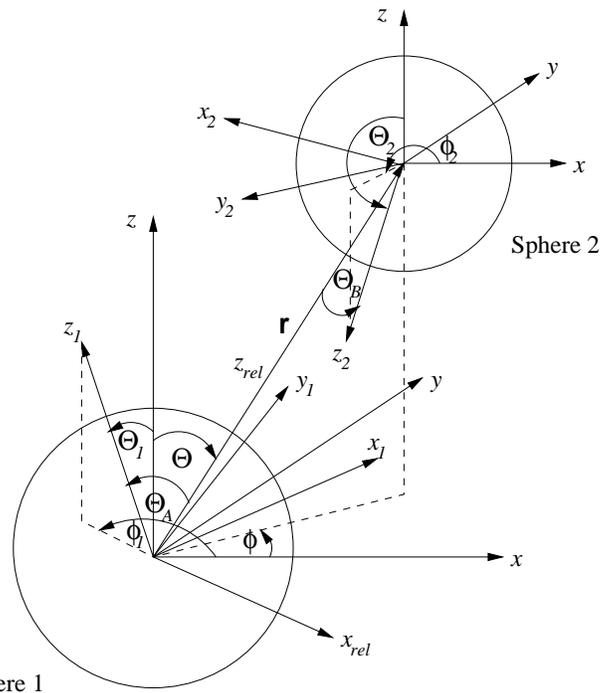}
\caption{\label{fig:model1} Absolute and relative coordinate system
describing the diffusional motion of the two spheres (see text). For
the sake of clarity, all $\chi$ and some of the $\phi$ angles have
been omitted in the drawing.}
\end{figure}

Our model for bimolecular association (see Fig.~\ref{fig:model1})
consists of two spherical molecules with radii $R_1$ and $R_2$,
respectively, whose relative distance and angular orientation change
by translational and rotational diffusion with diffusion constants
$D=D_1^{\text{trans}} + D_2^{\text{trans}}$, $D_1^{\text{rot}}$ and
$D_2^{\text{rot}}$.  The center of sphere 1 coincides with the origin
of a fixed-space coordinate system $\{x,y,z\}$. The position of the
center of sphere 2 is specified by the center-to-center vector \sfr\ 
whose spherical coordinates with respect to the fixed-space coordinate
system are given by $(r,\theta, \phi)$.

Each sphere carries a body-fixed coordinate system, denoted by
$\{x_1,y_1,z_1\}$ and $\{x_2,y_2,z_2\}$, respectively, with the axes
$z_1$ and $z_2$ pointing along \sfr\ when the two spheres are
perfectly aligned (and hence $z_1$ and $z_2$ can be thought of
pointing at the ``center'' of the reactive patch). The orientation of
these body-fixed coordinate systems with respect to the fixed-space
coordinate system $\{x,y,z\}$ is parametrized by sets of Euler angles
$\Psi_1=(\phi_1,\theta_1,\chi_1)$ and
$\Psi_2=(\phi_2,\theta_2,\chi_2)$. The angles $\phi_i$ and $\theta_i$,
$i=1,2$, are the usual azimuthal and polar coordinates of the $z_i$
axis, whereas $\chi_i$ measures the angle from the line of nodes,
defined to be the intersection of the $xy$ and the $x_iy_i$ planes, to
the $y_i$ axis. The set $(r,\theta, \phi,\Psi_1,\Psi_2)$ comprises the
absolute coordinates of the system.

For a convenient formulation of the reaction condition, we
additionally introduce a relative coordinate system
$\{x_{\text{rel}},y_{\text{rel}},z_{\text{rel}}\}$. The
$z_{\text{rel}}$ axis coincides with the center-to-center vector \sfr,
whereas the $x_{\text{rel}}$ axis lies in the plane spanned by \sfr\,
and the $z$ axis of the fixed-space coordinate system $\{x,y,z\}$. The
Euler angles $\Psi_A=(\phi_A,\theta_A,\chi_A)$ and
$\Psi_B=(\phi_B,\theta_B,\chi_B)$ specify the orientation of the
body-fixed coordinate systems $\{x_1,y_1,z_1\}$ and $\{x_2,y_2,z_2\}$
with respect to the coordinate system
$\{x_{\text{rel}},y_{\text{rel}},z_{\text{rel}}\}$.

\subsection{Reaction condition}

\begin{figure}
\includegraphics[scale=0.65]{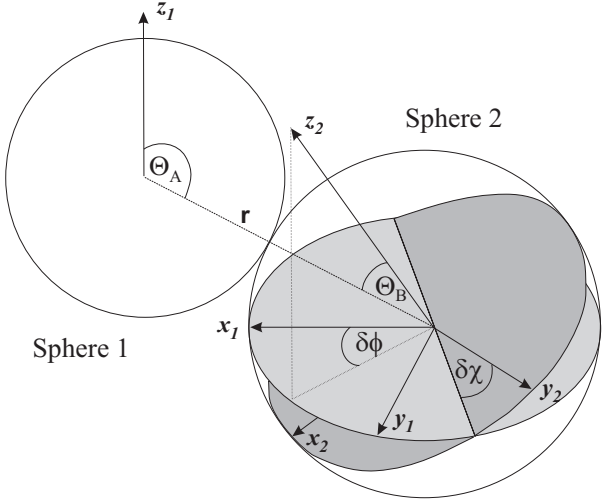}
\caption{\label{fig:model2}The axes and angles relevant to the reaction
  condition, Eqs.~\eqref{rc}. The angles $\theta_A$ and $\theta_B$
  measure how close the center of each reactive patch (coninciding
  with the body-fixed axes $z_1$ and $z_2$, respectively) is to
  the center-to-center vector \sfr.  The angles $\delta \phi$ and
  $\delta \chi$ denote relative torsion angles of the two body-fixed
  coordinate systems $(x_1, y_1, z_1)$ and $(x_2,y_2,z_2)$. To
  facilitate the visualization of these two angles, the origin of the
  $x_1$ and $y_1$ axes (belonging to the coordinate system of sphere
  1) has been shifted such as to coincide with the origin of the
  coordinate system of sphere 2.  Our reaction condition,
  Eqs.~\eqref{rc}, requires near-optimal alignment, \ie all angles
  $\theta_A$, $\theta_B$, $\delta \phi$, and $\delta \chi$ must be
  below given limits.}
\end{figure}

To fully specify the position and orientation of two rigid bodies,
nine variables are required, for instance, as introduced through our
absolute coordinate system, $(r,\theta, \phi,\Psi_1,\Psi_2)$. However,
for the expression of our reaction condition, only five variables,
describing the distance between the two spheres and their relative
orientation, are needed (see Fig.~\ref{fig:model2}). First, the
center-to-center distance is parametrized by $r$. The differences in
the orientation of the two spheres can be fully captured by the
differences in the Euler angles $\Psi_A=(\phi_A,\theta_A,\chi_A)$ and
$\Psi_B=(\phi_B,\theta_B,\chi_B)$, namely,
$\delta\theta=|\theta_A-\theta_B|$, $\delta\phi=|\phi_A-\phi_B|$, and
$\delta\chi=|\chi_A-\chi_B|$. Finally, we need a measure for the
extent to which the reactive patches on the spheres are aligned with
the center-to-center vector $\sfr$, which can be represented by the
sum of the polar angles $\theta_A+\theta_B$. To facilitate the
subsequent calculations, we replace the conditions on
$|\theta_A-\theta_B|$ and $\theta_A+\theta_B$ with independent
constraints on $\theta_A$ and $\theta_B$. Our reaction condition is
therefore
\be \label{rc}
\begin{cases}r = R_1 + R_2 \equiv
  R,\\
  \theta_{A,B} \le \theta_{A,B}^0,\\
  \delta\phi=|\phi_A-\phi_B| \le \delta\phi_0, \\
  \delta\chi=|\chi_A-\chi_B| \le \delta\chi_0. \end{cases}
\ee

\subsection{Derivation of the rate constant expression}

To derive an expression for the association rate constant, we
determine the concentration $\cc$ of spheres 2 as solution of the
steady-state translational--rotational diffusion equation in the
absolute coordinate system $\{r,\theta, \phi,\Psi_1,\Psi_2\}$
introduced in the preceding Sec.~\ref{sec:cs},
\bea\label{diffeq0}
\frac{\partial c}{\partial t} = 0 &=& \underbrace{D\nabla_{\sfr}^2\,
  c}_{\text{CoM motion}} + \underbrace{\Darot \biggl( \frac{\partial^2
    c}{\partial \delta_{x_1}^2} + \frac{\partial^2 c}{\partial
    \delta_{y_1}^2} + \frac{\partial^2 c}{\partial \delta_{z_1}^2}
  \biggr)}_{\text{rotation of sphere 1}} \nonumber \\ && +
\underbrace{\Dbrot \biggl( \frac{\partial^2 c}{\partial
    \delta_{x_2}^2} + \frac{\partial^2 c}{\partial \delta_{y_2}^2} +
  \frac{\partial^2 c}{\partial \delta_{z_2}^2}
  \biggr)}_{\text{rotation of sphere 2}}, \eea
where 
\be
\nabla_{\sfr}^2 = \frac{\partial^2}{\partial r^2}+\frac{2}{r}\frac{\partial
}{\partial r} + \frac{1}{r^2 \sin\theta}\frac{\partial}{\partial \theta}\biggl(
\sin\theta \frac{\partial}{\partial \theta} \biggr) +
\frac{1}{ r^2\sin^2\theta} \frac{\partial^2}{\partial \phi^2} 
\ee
is the Laplace operator acting on the center-to-center vector $\sfr$,
expressed in the spherical coordinates $(r,\theta,\phi)$, and the
$\delta_{s_i}$, $s=x,y,z$ and $i=1,2$, denote an infinitesimal
rotation of sphere $i$ about its body-fixed $s_i$ axis.
Equation~\eqref{diffeq0} can be viewed as composed of three individual
diffusional contributions, namely, the diffusional motion of the
center of mass of sphere 2 relative to sphere 1 and the rotational
diffusion of each sphere.

As in quantum mechanics, we can define angular
momentum operators $\hat{J}_{s_i} = -i\hbar\partial/\partial\delta_{s_i}$ as generators of infinitesimal rotations
of the spheres about their body-fixed axes, and can hence rewrite
Eq.~\eqref{diffeq0} as
\be\label{diffeq1}
0 = D\nabla_{\sfr}^2\, c + \Darot J^2_1 c + \Dbrot J^2_2 c, 
\ee
where $J^2_i=(-i\hbar)^{-2}
(\hat{J}_{x_i}^2+\hat{J}_{y_i}^2+\hat{J}_{z_i}^2)$. Using the
basic relations \cite{goldstein50}
\be
\begin{split}
d\delta_{x_i} &= d\theta_i \sin \chi_i - d\phi_i\sin\theta_i\cos\chi_i,\\
\quad d\delta_{y_i} &= d\theta_i \cos \chi_i + d\phi_i\sin\theta_i\sin\chi_i,\\
\quad d\delta_{z_i} &= d\phi_i \cos \theta_i + d\chi_i,
\end{split}
\ee
we can express the operator $J^2_i$ in terms of the
Euler angles $\Psi_i=(\phi_i,\theta_i,\chi_i)$,  
\begin{multline} \label{J}
J^2_i = \frac{1}{\sin\theta_i}\frac{\partial}{\partial \theta_i}\biggl(
\sin\theta_i \frac{\partial}{\partial \theta_i} \biggr) \\ +
\frac{1}{\sin^2\theta_i} \biggl( \frac{\partial^2}{\partial \phi_i^2}
+ \frac{\partial^2}{\partial \chi_i^2} - 2\cos\theta_i \frac{\partial^2}{\partial \phi_i\chi_i}\biggr).
\end{multline}
The advantage of the formulation of the diffusion equation,
Eq.~\eqref{diffeq0}, in terms of the operators $J^2_i$ in
Eq.~\eqref{diffeq1}, lies in the fact that the properties of the
$J^2_i$ are well-known, in particular their eigenfunctions, which are
given by the Wigner rotation matrices
$\mathcal{D}_{mn}^{l}(\phi,\theta,\chi)=e^{-im\phi}d_{mn}^l(\theta)e^{-in\chi}$
\cite{zare88}.

The general solution to Eq.~\eqref{diffeq1} that obeys the boundary
condition at $r\rightarrow \infty$,
\be\label{bc1}
\lim_{r\rightarrow \infty} \cc = c_0 = \text{const.},
\ee
can therefore be written as a series of products of
the eigenfunctions of $\nabla_{\sfr}^2$, $J^2_1$, and $J^2_2$, 
\begin{multline} \label{sol1}
\cc = c_0 + \sum_{ll_1l_2} \sum_{mm_1m_2} \sum_{n_1n_2} \alm \\ \times f_{ll_1l_2}(r)
Y_l^m(\theta,\phi) \mathcal{D}_{m_1n_1}^{l_1}(\Psi_1) \mathcal{D}_{m_2n_2}^{l_2}(\Psi_2), 
\end{multline}
where 
\be \label{bessel}
f_{ll_1l_2}(r) = \frac{K_{l+1/2}(\xi r)}{(\xi r)^{1/2}}
\ee
are the modified Bessel functions of the third kind \cite{arfken95}
(giving the desired behavior $f_{ll_1l_2}(r) \rightarrow 0$ as $r \rightarrow \infty$),
with $\xi \equiv [(\Darot/D) l_1(l_1+1) + (\Dbrot / D) l_2(l_2+1)]^{1/2}$.

For the boundary condition at $r=R$, the
usual, but analytically hardly tractable radiation boundary condition is
\be \label{radbc}
\frac{\partial c}{\partial r} \bigg\vert_{R} =
\frac{\kappa}{D} F(\Psi_A,\Psi_B) \ccR,
\ee
where $\kappa$ quantifies the extent of diffusion control in the
reaction, and $F(\Psi_A,\Psi_B) \equiv
\mathcal{H}(\theta_A^0-\theta_A)\mathcal{H}(\theta_B^0-\theta_B) \mathcal{H}(\delta
\phi_0-\delta \phi) \mathcal{H}(\delta
\chi_0-\delta \chi)$ represents the reaction condition
Eq.~\eqref{rc}, where $\mathcal{H}(x)$ is the step function
defined by $\mathcal{H}(x)=0$ for $x<0$ and $\mathcal{H}(x)=1$ for $x \ge
0$. 

In our approach, we express the radiation
boundary condition  using the constant-flux
approximation as introduced by Shoup
\etal \cite{shoup81}, by requiring that the flux is a
constant over the angular ranges in which the reaction can take place,
\be\label{bc2}
\frac{\partial c}{\partial r} \bigg\vert_{R} = QF(\Psi_A,\Psi_B),
\ee
and that Eq.~\eqref{radbc} is obeyed on the average over the surfaces
of the spheres, that is,
\begin{multline}\label{Q1}
\dall F(\Psi_A,\Psi_B) \, Q \\ = \frac{\kappa}{D} \dall
F(\Psi_A,\Psi_B) \, \ccR,
\end{multline}
where we have introduced the abbreviation $\int d\Omega \equiv \int
\sin\theta d\theta \, \int d\phi$. 

To proceed, we express $F(\Psi_A,\Psi_B)$ in
absolute coordinates. First, we expand $F(\Psi_A,\Psi_B)$ in terms of
rotation matrices,
\bea \label{F}
F(\Psi_A,\Psi_B) &=& \sum_{l_Al_B}\sum_{m_An_A}\sum_{m_Bn_B}
C_{l_Al_B}^{m_An_Am_Bn_B} \nonumber \\ &\times& \mathcal{D}_{m_An_A}^{l_A}(\Psi_A)
\mathcal{D}_{m_Bn_B}^{l_B}(\Psi_B),
\eea
where the expansion coefficients $C_{l_Al_B}^{m_An_Am_Bn_B}$ are given
by 
\bea \label{c0} 
C_{l_Al_B}^{m_An_Am_Bn_B} &=& \frac{2l_A+1}{8\pi^2}
\frac{2l_B+1}{8\pi^2} \int d\Psi_A \, \int d\Psi_B \nonumber \\ &
\times & \mathcal{D}_{m_An_A}^{l_A*}(\Psi_A)
\mathcal{D}_{m_Bn_B}^{l_B*}(\Psi_B) F(\Psi_A,\Psi_B) \nonumber \\
&=& \frac{2l_A+1}{8\pi^2} \frac{2l_B+1}{8\pi^2} \nonumber \\ & \times
& \frac{4\pi\sin(m_A\delta\phi_0)}{m_A}
\frac{4\pi\sin(n_A\delta\chi_0)}{n_A}
\nonumber \\
&\times & \int_0^{\theta_A^0} \sin\theta_A d\theta_A\,
d^{l_A}_{m_An_A}(\theta_A) \nonumber \\ & \times & \int_0^{\theta_B^0}
\sin\theta_B d\theta_B\, d^{l_B}_{-m_A-n_A}(\theta_B)
\nonumber \\
&\equiv& \frac{2l_A+1}{8\pi^2} \frac{2l_B+1}{8\pi^2}
\widehat{C}_{l_Al_B}^{m_An_A}.  
\eea
The absolute coordinate system $\{x,y,z\}$ can
be transformed into the relative coordinate system
$\{x_{\text{rel}},y_{\text{rel}},z_{\text{rel}}\}$ by rotations
through the three Euler angles $(\phi-\pi,\theta,0)$. The
corresponding transformations of the rotation matrices appearing in
Eq.~\eqref{F} are then
\bea \label{trsf}
\mathcal{D}_{m_An_A}^{l_A}(\Psi_A) =
\sum_{m_1} \mathcal{D}_{m_1m_A}^{l_A}(\phi-\pi,\theta,0)
\mathcal{D}_{m_1n_A}^{l_A}(\Psi_1), \nonumber \\
\mathcal{D}_{m_Bn_B}^{l_B}(\Psi_B) =
\sum_{m_2} \mathcal{D}_{m_2m_B}^{l_B}(\phi-\pi,\theta,0)
\mathcal{D}_{m_2n_B}^{l_B}(\Psi_2).\nonumber  
\eea
The expansion coefficients $\alm$ in Eq.~\eqref{sol1} can be obtained
by substituting the expansion for $F(\Psi_A,\Psi_B)$, Eq.~\eqref{F},
expressed in absolute coordinates $(r,\theta, \phi,\Psi_1,\Psi_2)$,
using the above transformations, into Eq.~\eqref{bc2}, which yields
\bea \label{alm} 
\alm &=& \frac{Q}{f'_{ll_1l_2}(R)}
(-1)^{m+m_1+m_2-n_1-n_2} \nonumber \\ & \times & \sqrt{4\pi(2l+1)}
\bigl( \begin{smallmatrix} l & l_1 & l_2 \\ m & -m_1 & -m_2
\end{smallmatrix} \bigr) \nonumber \\ & \times & \sum_{m_A}
\widehat{C}_{l_1l_2}^{m_A-n_1} \bigl( \begin{smallmatrix} l & l_1 &
  l_2 \\ 0 & m_A & -m_A \end{smallmatrix} \bigr),
\eea
where $\bigl( \begin{smallmatrix} l&l_1&l_2\\m&m_1&m_2
\end{smallmatrix} \bigr)$ is the Wigner 3-$j$ symbol. Evaluating
Eq.~\eqref{Q1} using the expansion coefficients, Eq.~\eqref{alm}, yields for the constant $Q$    
\bea \label{Q} %
Q &=& c_0a_o \times \Biggl[ \frac{D}{\kappa}a_0 - \sum_{ll_1l_2}
\frac{f_{ll_1l_2}(R)}{f'_{ll_1l_2}(R)} 4\pi (2l+1) \nonumber \\ &
\times & \frac{2l_1+1}{8\pi^2} \frac{2l_2+1}{8\pi^2}
\sum_{n=-l_1}^{+l_1} \bigl[ \sum_{m=-l_1}^{+l_1}
\widehat{C}_{l_1l_2}^{mn} \bigl( \begin{smallmatrix} l & l_1 & l_2 \\
  0 & m & -m \end{smallmatrix} \bigr) \bigr]^2 \Biggr]^{-1} 
\eea
where we have introduced 
\bea 
a_0 &=& \dall F(\Psi_A,\Psi_B) \nonumber \\ & = & (4\pi)^3
\delta\phi_0 \delta\chi_0 (1-\cos\theta_A^0) (1-\cos\theta_B^0). 
\eea 
$a_0/(4\pi\times 8\pi^2 \times 8\pi^2)$ is the fraction of angular
orientational space over which the reaction can occur. In deriving
Eqs.~\eqref{alm} and \eqref{Q}, we have made use of the identities \cite{zare88}
\begin{gather*}
  Y_l^{m*}(\theta,\phi) = \sqrt{\frac{2l+1}{4\pi}}
  \mathcal{D}_{m0}^{l}(\phi,\theta,0),  \\
  \int d\Psi \mathcal{D}_{m_1n_1}^{l_1*}(\Psi)
  \mathcal{D}_{m_2n_2}^{l_2}(\Psi) = \frac{8\pi^2}{2l_1+1}
  \delta_{l_1l_2} \delta_{m_1m_2} \delta_{n_1n_2},
  \\
  \int d\Psi \mathcal{D}_{m_1n_1}^{l_1}(\Psi)
  \mathcal{D}_{m_2n_2}^{l_2}(\Psi)
  \mathcal{D}_{m_3n_3}^{l_3}(\Psi) \phantom{hspace{1.3cm}} \\
  \phantom{hspace{1.3cm}} = 8\pi^2 \bigl( \begin{smallmatrix} l_1 &
    l_2 & l_3 \\ m_1 & m_2 & m_3
\end{smallmatrix} \bigr) \bigl( \begin{smallmatrix}  l_1 & l_2 & l_3
  \\ n_1 & n_2 & n_3
\end{smallmatrix} \bigr), \\
\sum_{m_1m_2} \bigl( \begin{smallmatrix} l_1 & l_2 & l_3 \\ m_1 & m_2
  & m_3
\end{smallmatrix} \bigr) \bigl( \begin{smallmatrix}  l_1 & l_2 & l'_3
  \\ m_1 & m_2 & m'_3 
\end{smallmatrix} \bigr) =
\frac{1}{2l_3+1}\delta_{l^{\phantom{\prime}}_3l^{\prime}_3}
\delta_{m^{\phantom{\prime}}_3m^{\prime}_3}.
\end{gather*}
The diffusion-controlled rate constant is given by 
\bea \label{k0} 
\kDC &=& \frac{1}{(8\pi^2)^2} \frac{R^2 D}{c_0} \dall
\frac{\partial c}{\partial r} \bigg\vert_{R} \nonumber \\ &=&
\frac{1}{(8\pi^2)^2} \frac{R^2 D}{c_0} a_0 Q.
\eea
Since the functions $f_{ll_1l_2}(r)$, defined in Eq.~\eqref{bessel}, obey the recursion relation
\be \label{recurs} 
f^{\prime}_{ll_1l_2}(r) = \frac{l}{r}f_{ll_1l_2}(r)
- \xi f_{(l+1)l_1l_2}(r), 
\ee
the final expression for the diffusion-limited rate constant,
Eq.~\eqref{k0}, becomes
\bea \label{k1} 
\kDC &=& D (Ra_0/8\pi^2)^2 \times \Biggl[
\frac{D}{\kappa}a_0 \nonumber \\ && - R \sum_{ll_1l_2}
\frac{K_{l+1/2}(\xi^*)}{l K_{l+1/2}(\xi^*) - \xi^* K_{l+3/2}(\xi^*)}
\nonumber \\ &\times& 4\pi (2l+1) \frac{2l_1+1}{8\pi^2}
\frac{2l_2+1}{8\pi^2} \nonumber \\ & \times & \sum_{n=-l_1}^{+l_1}
\biggl( \sum_{m=-l_1}^{+l_1} \widehat{C}_{l_1l_2}^{mn} \, \bigl(
\begin{smallmatrix} l & l_1 & l_2 \\ 0 & m & -m
\end{smallmatrix} \bigr) \biggr)^2 \, \Biggr]^{-1},
\eea
with $\xi^* = \xi R$.

\section{Results} \label{sec:results}

\subsection{Limiting cases} \label{sec:limits}

\subsubsection{Axially symmetric reactive patches} \label{sec:repss}

Zhou \cite{zhou93} presented an analytical expression for the
association rate constant of two spherical molecules bearing axially
symmetric patches. In the notation of our model, this corresponds to
setting $\delta \phi_0 = \delta\chi_0 = \pi$, which makes
$\widehat{C}_{l_1l_2}^{mn}=0$ in Eqs.~\eqref{c0} and \eqref{k1},
unless $m=n=0$. Using
$\mathcal{D}_{00}^l(\phi,\theta,\chi)=d_{00}^l(\phi,\theta,\chi)=P_l(\cos\theta)$,
where $P_l(\cos\theta)$ are the Legendre polynomials, the expression
for the rate constant, Eq.~\eqref{k1}, becomes
\bea \label{k-zhou}
\kDC &=& 4 \pi D R^2 (1-\cos\theta_A^0)^2 (1-\cos\theta_B^0)^2
\nonumber \\ &\times& \bigg[
4\frac{D}{\kappa} (1-\cos\theta_A^0) (1-\cos\theta_B^0) \nonumber \\
&& - R\sum_{ll_1l_2} 
\frac{K_{l+1/2}(\xi^*)}{l K_{l+1/2}(\xi^*) - \xi^* K_{l+3/2}(\xi^*)}
\nonumber \\ &\times& (2l+1) (2l_1+1) (2l_2+1) \nonumber \\
& \times & \bigg( \int_0^{\theta_A^0} \sin\theta_A
d\theta_A\, P_{l_1}(\cos\theta_A) \bigg)^2
\nonumber \\ &\times& \bigg(\int_0^{\theta_B^0} \sin\theta_B d\theta_B\,
P_{l_B}(\cos\theta_B) \bigg)^2 \bigl( \begin{smallmatrix} l & l_1 & l_2 \\ 0 & 0 & 0 
\end{smallmatrix} \bigr)^2 \bigg]^{-1}
\eea
which agrees with the solution presented by Zhou \cite{zhou93}.

\subsubsection{Uniform reactivity}

If we assume that one sphere is uniformly reactive and the other has
an axially symmetric patch (that is, $\delta \phi_0 = \delta\chi_0 =
\pi$ and $\theta_B^0=\pi$), we arrive at the model introduced by
\v{S}olc and Stockmayer \cite{solc71}.  Then, since $\theta_B^0 = \pi$
and $\int_{0}^{\pi} \sin\theta d\theta \, P_{l}(\cos\theta) = 0$ if $l
\not= 0$, only the term $l_2=0$ (and hence $l=l_1$) gives a nonzero
contribution to the sum in Eq.~\eqref{k-zhou}. Then Eq.~\eqref{k-zhou}
reduces to
\bea \label{k-ss}
\kDC &=& 2 \pi D R^2 (1-\cos\theta_A^0)^2 \times \bigg[
\frac{D}{\kappa} (1-\cos\theta_A^0)  \nonumber \\
&& - R\sum_l
\frac{(l+1/2) K_{l+1/2}(\xi^*)}{l K_{l+1/2}(\xi^*) - \xi^* K_{l+3/2}(\xi^*)}
\nonumber \\ &\times&  \bigg( \int_0^{\theta_A^0} \sin\theta_A
d\theta_A\, P_{l_1}(\cos\theta_A) \bigg)^2 \bigg]^{-1},
\eea
where now $\xi^*=R[(\Darot/D) l(l+1)]^{1/2}$, which coincides with the result of \v{S}olc and Stockmayer
\cite{solc73} and Shoup \etal \cite{shoup81}.

Assuming both spheres to be uniformly reactive,
$\theta_A^0=\theta_B^0=\delta \phi_0 = \delta\chi_0 = \pi$, only the
term $l=l_1=l_2=0$ contributes, and hence $\xi^* = 0$. Because
$K_{1/2}(\xi^*)/\xi^* K_{3/2}(\xi^*) \rightarrow 1$ as $\xi^*
\rightarrow 0$, Eq.~\eqref{k-ss} becomes, in the fully
diffusion-controlled case ($\kappa \rightarrow \infty$), $\kDC = 4\pi D
R$, which is just the classic Smoluchowski diffusion-controlled rate
constant for two uniformly reactive spheres.

\subsection{Numerical evaluation} \label{sec:values}

In the following, we shall assume the reaction to be fully
diffusion-controlled ($\kappa \rightarrow \infty$), and take the radii
of the two spheres to be identical, $R_1=R_2$.  Instead of plotting
the absolute value of the association rate constant $\kDC$, we
introduce the dimensionless relative association rate constant $\kDC^*
= \kDC / 4\pi D R$, which is the ratio of the orientation-constrained
rate constant to the Smoluchowski rate constant for two uniformly
reactive spheres.

\begin{figure}
\input{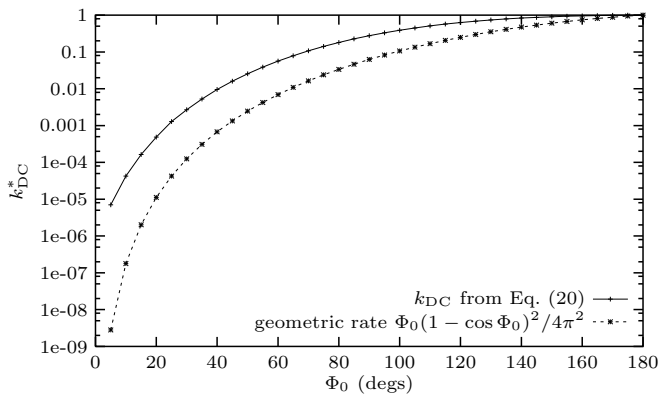}
\caption{\label{fig:rates} Diffusion-controlled ($\kappa \rightarrow
  \infty$) relative association rate constant $\kDC^*=\kDC/4\pi D R$,
  with $\kDC$ computed from Eq.~\eqref{k1}, as a function of the
  angular constraint $\Phi_0 \equiv \theta_A^0=\theta_B^0=\delta
  \phi_0 = \delta\chi_0$ (solid curve). Also shown is the rate
  expected from a simple probabilistic
  argument, $\kDC=\Phi_0 (1-\cos \Phi_0)^2/4\pi^2$ (geometric rate;
  dashed curve).} 
\end{figure}

The full dependence of the relative association rate constant on
$\theta_A^0$, $\theta_B^0$, $\delta\phi_0$, and $\delta\chi_0$ is not
easy to display in a single plot. For simplicity, we set all four
parameters equal, and in Fig.~\ref{fig:rates} plot the relative
association rate constant $\kDC^*$ computed from Eq.~\eqref{k1} as a
function of a single parameter (referred to $\Phi_0$ in the
following). For comparison, we also show the relative association rate
expected from a purely probabilistic argument (geometric rate), given
by the fraction of angular orientational space over which the reaction
can occur, $a_0/(4\pi\times 8\pi^2 \times 8\pi^2)=\Phi_0 (1-\cos
\Phi_0)^2/4\pi^2$.

It is evident from Fig.~\ref{fig:rates} that the difference between
the rate constant $\kDC^*$ and the geometric rate gets more striking
as the angular constraint $\Phi_0$ becomes more stringent. For
instance, in the important case of sterically highly specific
protein--protein interactions where $\Phi_0$ will typically range
between $5^{\text{o}}$ and 15$^{\text{o}}$, the geometric rate is
about 2 to 3 orders of magnitude too low, as compared with the
association rate computed from Eq.~\eqref{k1}.

\subsection{\label{sec:bd}Comparison against Brownian dynamics simulations} 

In the Brownian dynamics simulations by Northrup and Erickson
\cite{northrup92}, protein molecules are modeled as hard spheres of
$R=18\text{~\AA}$ diffusing in water ($\eta \simeq 8.9\times
10^{-4}\, \text{Ns/m$^2$}$) at $T=298$ K; no forces are assumed to act
between the molecules. The translational and rotational diffusion
constants are computed from the Stokes--Einstein relations
$D^{\text{trans}}=k_BT/6\pi \eta R$ and $D^{\text{rot}}=k_BT/8\pi \eta
R^3$, respectively.

Instead of angular constraints, the model uses a contact-based
reaction condition. A set of four distinctly numbered contact points
is mounted on each sphere in a $17\text{~\AA} \times 17\text{~\AA}$
square arrangement on a plane tangential to the surface of the sphere.
Reaction is assumed to occur when at least three of the four contact
points are correctly matched and within a maximum distance of 2~\AA.

We performed numerical simulations to estimate the angles
$\theta_A^0$, $\theta_B^0$, $\delta \phi_0$, and $\delta \chi_0$ (as
defined in our model, see Sec.~\ref{sec:theory}) that correspond to
this contact-based reaction condition. Clearly, there will be a
multiplicity of sets of these angles for which the contact-based
reaction criterion is met. To reduce the search space in a reasonable
way, we looked for geometric configurations where all four angles were
equal, $\theta_A^0=\theta_B^0=\delta \phi_0=\delta \chi_0$, and found
that the contact-based reaction condition can be well represented by
an angular constraint of $\theta_A^0=\theta_B^0=\delta \phi_0=\delta
\chi_0=6.7^{\text{o}}$.

With these angular constraints, numerical evaluation of Eq.~\eqref{k1}
with the parameters specified above and $\kappa \rightarrow \infty$
gives $\kDC=1.04 \times 10^{5}\, \text{M$^{-1}$ s$^{-1}$}$, which is
in very good agreement with the value obtained from the Brownian
dynamics simulation by Northrup and Erickson \cite{northrup92}, $\kDC
= 1 \times 10^{5} \, \text{M$^{-1}$ s$^{-1}$}$.

\section{Summary}

We have presented a general expression for the diffusion-controlled
association rate of two molecules where reaction can occur solely if
specified constraints on the mutual orientation are fulfilled. Our
solution goes far beyond previous treatments in the ability to impose
very general, asymmetric orientational constraints, as needed for
instance in a proper description of the sterically highly specific
association of two proteins.

Since our expression for the rate constant, Eq.~\eqref{k1}, was
derived under the assumption of no forces acting between the two
molecules, a comparison of measured association rates with their
theoretical values calculated from Eq.~\eqref{k1} should reveal the
extent to which long-range interactions contribute to the rate of
intermolecular association. Such an investigation would be of
particular interest in the case of the association of proteins with
small ligands and other proteins.

\begin{acknowledgments}
  This work was supported by a grant from the National Institute of
  Health.
\end{acknowledgments}

\end{document}